# Crowd-Powered Sensing and Actuation in Smart Cities: Current Issues and Future Directions

Jiangtao Wang, Yasha Wang, Daqing Zhang, Qin Lv, and Chao Chen

**Abstract**: With the advent of seamless connection of human, machine, and smart things, there is an emerging trend to leverage the power of crowds (e.g., citizens, mobile devices, and smart things) to monitor what is happening in a city, understand how the city is evolving, and further take actions to enable better quality of life, which is referred to as *Crowd-Powered Smart City* (*CPSC*). In this article, we provide a literature review for CPSC and identify future research opportunities. Specifically, we first define the concepts with typical CPSC applications. Then, we present the main characteristics of CPSC and further highlight the research issues. In the end, we point out existing limitations which can inform and guide future research directions.

**Keywords**: crowd-powered computing, smart city, context sensing and actuation.

## 1. Introduction

To tackle the challenges in providing good quality of life for citizens (e.g., environmental protection, designing efficient public transit systems, population control, etc.), the concept of "smart city" emerges by integrating information and communication technology (ICT) and Internet of things (IoT) [1]. Smart city solutions allow us to monitor what is happening in a city, how the city is evolving, and how to change the running process of the city to enable better quality of life.

In this article, we classify the system functionalities of smart cities into the following two categories from the perspective of context-aware computing: (1) *Urban Context Sensing*. The goal of urban context sensing is to collect real-time information in a city, including environmental information (e.g., air quality, noise), infrastructure status (e.g., missing manholes, broken streetlights), and city dynamics (e.g., traffic congestions, flow of people), etc. With urban context sensing, we can have a better understanding of the current urban context, which helps us better monitor and manage the city and society. (2) *Urban Context Actuation*. While urban context sensing aims to obtain information in a city, the goal of urban context actuation is to impose actions or influence on the urban context, thus changing how the city will be managed. Urban context actuation tries to optimize different smart city systems (such as intelligent transportation systems, smart logistic systems, etc.) by adding, adjusting, and transferring the limited resources (e.g., deploying new bus routes and stations, assigning vehicles to deliver packages, etc.).

Traditional solutions for smart cities follow the "god-dominant" paradigm, in which powerful organizations or persons (just like the god) have full control in designing how urban context sensing and actuation should be executed. For urban context sensing, traditional systems usually rely on specialized infrastructure (e.g., air quality monitoring stations, surveillance cameras), which requires a high cost for deployment and maintenance. For urban context actuation, powerful organizations or enterprises determine when, where, and how to deploy or re-distribute different types of resources (e.g., constructing new pavement, rebalancing shared bicycles). As the decision making process highly depends on the domain knowledge of experts, such god-dominant urban context actuation may not be efficient and scientific in some complicated scenarios.

In recent years, the dramatic technology progress in mobile/wearable computing, IoT, and cloud computing has enabled seamless connection of the cyber and physical space in a city, which makes the hybrid computing of human, machine, and smart things a new trend. Given this background, there is now a great opportunity to integrate the power of crowds (e.g., citizens, mobile devices, and smart things) into various urban context sensing and actuation tasks, which is complementary to the traditional god-dominant solutions. In this article, we refer to this emerging smart city computing paradigm as *Crowd-Powered Smart City* (*CPSC*), which consists of the following two aspects.

- *Crowd-Powered Urban Context Sensing* (*CPUCS*). With the prevalence of sensor-rich smartphones, CPUCS (including mobile crowdsensing, participatory sensing, and human-centric sensing [2]) has become a new sensing paradigm. CPUCS leverages the mobility of mobile users, the sensors embedded in mobile phones, and existing communication infrastructure to fulfill urban context sensing tasks. Compared with traditional infrastructure-based approaches, crowd-powered urban context sensing can sense large urban regions with less cost and higher efficiency.

- *Crowd-Powered Urban Context Actuation* (*CPUCA*). Ordinary citizens can work collaboratively to complete various kinds of tasks (e.g., rebalancing shared bikes, package delivery, and travel route recommendations), or be directed to act in a more coordinated fashion (e.g., vehicle-to-vehicle collaboration in driving, sharing reserved tables, etc.). Compared with the god-dominant paradigm, CPUCA leverages the wisdom of the crowds to optimize the efficiency of smart city systems.

In this article, we focus specifically on the state-of-the-art research works for CPSC and provide a tutorial with future research opportunities. We first define the basic concepts in CPSC and compare its research scope with other relevant research topics (Section 2). Then, we present the applications and main characteristics of CPSC (Section 3 and 4). We further present the research issues with corresponding techniques (Section 5). Finally, we point out the limitations to inspire and guide future research directions (Section 6) and conclude the article (Section 7).

## 2. Concepts and Research Scope

In this section, we first define the basic concepts in CPSC with two running examples, and then compare the research scope with some overlapping research topics.

### 2.1 Concepts Definition

In this article, CPSC (Crowd-Powered Smart City) is formally defined as *the computing paradigm in smart cities, where the power of crowds (including human, mobile devices, and smart things, etc.) are utilized to monitor what is happening in a city, understand how the city is evolving, and further take actions to enable better quality of life*. CPSC consists of two major aspects, which are urban context sensing and actuation with crowds (CPUCS and CPUCA). Here, we present two running examples of CPSC, which will be used throughout the article. One example is for CPUCS (Example 1), and the other one is for CPUCA (Example 2).

(**Example 1: Air Quality Monitoring**) *AirSense is a crowd-powered urban context sensing application,*

*which continuously provides real-time air quality information (see Fig 1). A large number of mobile users serve as the "human sensor" to collect the air quality information in each spatial-temporal cell using their mobile devices. By aggregating the sensing data from the crowds, we can obtain a real-time and city-scale air quality map*

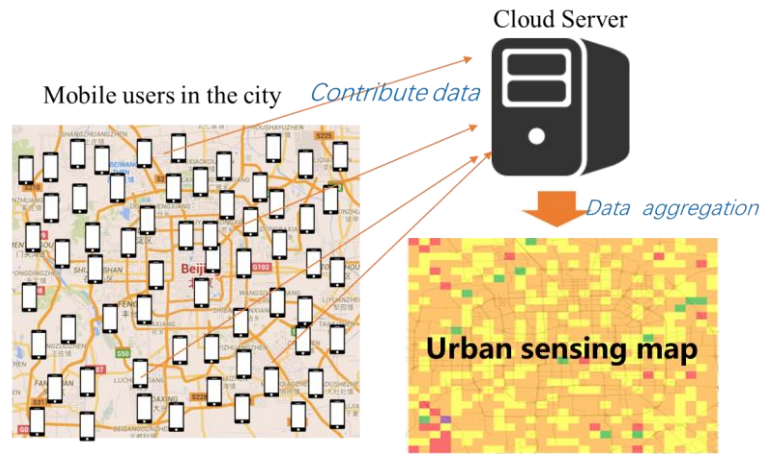

Fig. 1 AirSense: An example of CPUCS, in which mobile users are assigned air quality sensing tasks for different subareas and time periods.

(**Example 2: Rebalancing Shared Bikes**) *Bike sharing systems (BSS) are very popular in modern cities nowadays. Due to the imbalance between the renting and return rates, operators need to re-distribute the bikes among different stations. Employing dedicated staff and vehicles to balance the system incurs large operational costs and undermines the green concept of BSS. Alternatively, we can use the idea of CPUCA to address this issue, where we engage the users of such systems and provide incentives for them to re-balance the system. As illustrated in Fig. 2, if the nearest bike station is almost empty, users are encouraged to walk to an alternative station to rent bikes with incentive rewards. Similarly, if the nearest station is almost full, we can encourage users to return their bikes to alternative stations.*

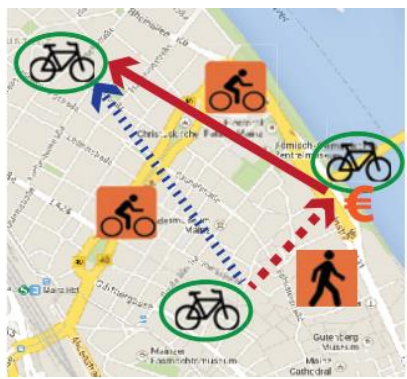

Fig. 2 Rebalancing shared bike: An example of CPUCA, in which participants are given incentive rewards to rent or return bikes from alternative stations instead of the nearest station.

To enable the new computing paradigm of CPSC, the basic elements in CPSC are defined as follows, and Fig. 3 provides a summary of these elements and their relationships.

- *Urban context* is the status of objects (e.g., city infrastructure) or phenomenon (e.g., air quality) in a city. In the example of air quality sensing, the urban context is the air quality readings at different subareas and time periods. For the bike rebalancing scenario, the urban context is the status of different bike stations (e.g., number of bikes).

- *CPSC task* is defined as a task to sense or change the urban context. The tasks which aim to sense (or "read") the urban context are called CPUCS tasks, while the tasks whose goal is to change (or "write") the urban context are defined as CPUCA tasks.

- *Participants* are ordinary citizens or smart things (e.g., wearable devices, smart mobile phones, and smart cars) who/which can work collaboratively to complete CPSC tasks. There are two types of roles for the participants. Participants of CPUCS tasks are called *crowd sensors*, while participants of CPUCA tasks are called *crowd actuators*.

- *Task Organizers* are those who are responsible for managing and coordinating CPSC tasks. For example, the task organizer can be the city government in the air quality monitoring task, or the operator of bike sharing systems in the bike rebalancing task.

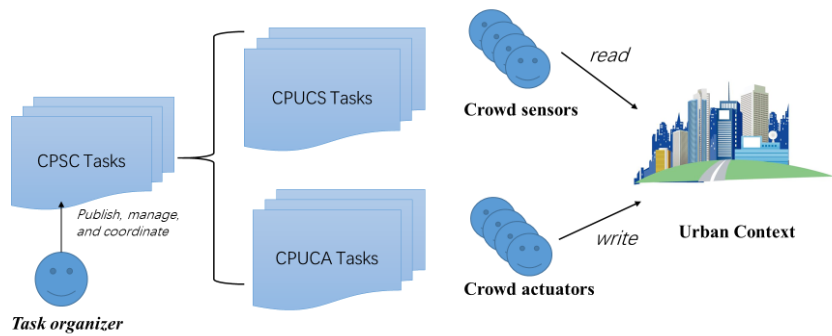

Fig. 3 Basic concepts, elements and their relationships in CPSC.

## 2.2 Comparison of Scope with Overlapping Topics

CPSC is related to several other research topics. Here, we list these other topics and discuss their research scopes and relationships with CPSC. Fig. 4 illustrates the relationship among CPSC and other topics.

- *Mobile Crowdsensing/Participatory Sensing*. Mobile Crowdsensing (MCS) or participatory sensing [2] leverages sensor-rich mobile devices of a large number of participants and their inherent mobility to obtain an aggregated picture of certain phenomenon (e.g., air quality or noise map). Compared with CPSC, the scope of MCS or participatory sensing is narrower as it focuses only on information collection in a city without involving urban context actuation.

- *Crowdsourcing*. The term "crowdsourcing" was proposed by Jeff Howe at Wired [3], to describe how businesses were using the Internet to "outsource work to the crowd". The similarity between crowdsourcing and CPSC is that both leverage the power of the crowd to complete complex tasks. However, crowdsourcing mainly focuses on online crowd work in the cyber space (e.g., crowdsourcing-based software engineering and crowdfunding), while CPSC primarily focuses on sensing or actuation in the physical space (e.g., physical objects or phenomenon in a city).

- *Urban Computing*. Urban computing is a process of collecting, fusion, and analysis of big and heterogeneous data generated from diverse sources in a city [4]. The core goal of urban computing is to extract intelligence from big data for better understanding of a city, which falls mainly into the knowledge discovery and data mining community. Urban computing overlaps with CPSC in data collection and analysis, but it does not include the aspect of urban context actuation in CPSC.

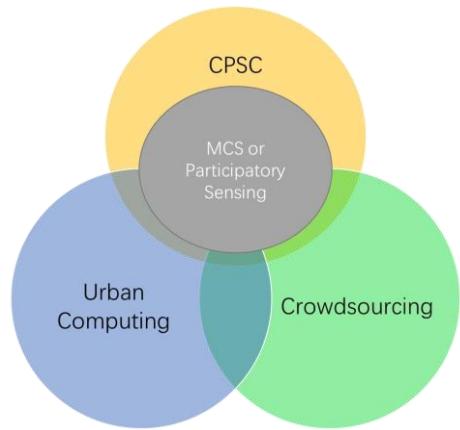

Fig. 4 An illustration of the relationships among CPSC and other research topics with similar or overlapping concepts.

## 3. CPSC Applications

In this section, we present some typical CPSC applications, including both CPUCS and CPUCA applications.

### 3.1 CPUCS Applications

The development of CPUCS has resulted in various novel applications, which can be divided into three categories: environment sensing, infrastructure sensing, and social sensing applications.

- *Environment Sensing*. This type of applications aims at collecting environmental information in a city (e.g., air quality, noise level, crowd density, etc.). For example, Ear-phone [5] provides a noise sensing map in the city based on the idea of participatory sensing. SakuraSensor [6] leverages crowd-sourced video stream to highlight beautiful cherry-lined roads.

- *Infrastructure Sensing*. Representative Studies include traffic congestion detection, place characterization and parking space availability detection. For example, the authors in [7] proposed an approach for city-scale traffic status estimation with probe vehicles. Another typical example application is ParkNet [8], where vehicles collect parking space occupancy information and a real-time map of parking availability is obtained.

- *Social Sensing*. This set of applications attempt to sense citizens' social aspects through crowd

sensing. For instance, SociableSense [9] is a smartphone-based platform for sensing social relations and interactions among users, which provides the users with a quantitative measure of their sociability and cooperation efficiency and that of their colleagues.

### 3.2 CPUCA Applications

There are various CPUCA applications which aim at changing the urban context, and we present some typical studies as follows.

- *Rebalancing Shared Bikes*. The user incentive based approaches utilize the crowd-based mechanism to incentivize users in the bike rent-and-return process by providing monetary incentives. For example, the authors in [10] design different incentive mechanisms and task assignment optimization policies, which encourage users to drop-off at nearby stations with insufficient bikes and pick-up at entirely-full stations.

- *Urban Crowd Steering*. In crowd steering [11], collective movements of people are tracked and movements are encouraged via mobile app advice or other mechanisms. The reasons for steering the crowd might be for emergency evacuation, guided tours, safe movement of people during large rallies and concerts, regulating the use of spaces or for commercial purposes (e.g., steer crowds to move through certain businesses areas).

- *Mobile Crowdsourcing based Logistics.* In the area of logistics, a specific CPUCA instance is when people go about their daily lives, they have the opportunity to carry packages and deliver to specific locations or individuals. For example, the authors in [12] study the package delivery tasks through mobile crowdsourcing. It evaluates the performance using mobility trace dataset, and the results show that packages can be delivered with required speed and coverage.

- *Vehicle-to-Vehicle Coordination*. When vehicles know which roads are congested, the navigation system will try to recommend better routes. But if all vehicles were to do so, the less congested route would also become congested. Motivated by this, the studies, such as [13], exploit how to improve the cooperation among cars in using different routes to achieve a optimal overall utility.

- *Resource Sharing*. Resource sharing among the crowds can optimize the resource utilization in a city. A platform for connecting consumers and providers can be effective, from sharing restaurant reservations [15] to car parking spaces [14].

### 4. CPSC Characteristics

In this section, we present the main characteristics of CPSC which differentiate it from traditional god-dominant smart city computing.

- **Computation resources: Human, machine, and mixed**. From the perspective of computation resources, the power of crowds utilized in CPSC can be divided into three classes.1) *Machine computation*: the crowd of machine resources used to complete CPSC tasks can range from CPU to storage, GPS access, Internet connectivity, and the sensing capabilities. For example, a large number of smartphones may automatically collect and calculate the AQI (air quality index) of the surrounding environment without the participation of human. 2) *Human computation*: the human participants perform CPSC tasks manually. For example, in the crowdsourcing-based logistic [12], human participants carry packages and deliver to specific locations or individuals. 3) *Mixed human-machine computation*: participants and computers can work together to complete CPSC tasks. For example, a car with the permission of its owner may allow another car to send communication messages.

- **Human participation: opportunistic or participatory**. Human participants can participate in CPSC tasks in two different modes, the opportunistic mode and the participatory mode. (1) In the opportunistic mode, the participants complete CPSC tasks during their daily routines without the need to change their routes (e.g., the air quality monitoring task). (2) In the participatory mode, however, participants are required to change their original routes and move specifically to certain locations (e.g., the rebalancing shared bikes task). The opportunistic mode does not require knowledge of the participants' intended travel routes, so it is less intrusive for the participants and less costly for the task organizers. However, the completion of tasks depends heavily on the participants' routine trajectories. Tasks that are located at places visited by few or even no participants are less likely to be completed. The participatory mode requires participants to move specifically to task locations, which can guarantee task completion. However, since participants need to deviate from their original routines and travel to task locations, it incurs extra travel cost and can be intrusive.

- **Participant-side factors**. For the success of CPSC, it is crucial to attract a large number of participants. However, there are several factors that significantly affect people's willingness to participate. These factors can be divided into two categories: *concerns and motivations*. (1) *Concerns* are issues that may reduce the participation willingness, such as intrusiveness, smartphone energy consumption, mobile data cost, and privacy leaking risk. 2) *Motivations* are incentive mechanisms that encourage participation. For example, financial or monetary gain has been an important incentive method. In addition, people could be motivated to participate in an activity by social and ethical reasons. With the above two types of factors in mind, we can make efforts in either addressing these concerns (e.g., designing less-intrusive computer-human interfaces, energy-saving techniques and privacy preserving mechanisms) or designing appropriate incentive mechanisms.

## 5. Major Research Issues and Solutions

The research issues in different CPSC tasks can be very different. In this section, we present the major and common research issues among different types of CPSC tasks and corresponding solutions.

- **Participant selection and task assignment**. Automatic participant selection or task assignment is crucial for the task completion quality and efficiency of CPSC. With different goals and constraints, participant selection or task assignment in CPSC can be formulated as combinatorial optimization problems, which attempt to find an optimal solution from a large search space. For urban context sensing tasks, the main factors of optimization include sensing quality and reliability, spatial-temporal coverage, energy consumption, incentive budget, etc. For urban context actuation tasks, the optimization factors mainly include quality of service, incentive budget, human intrusiveness, etc. Intuitively, it is easy to think of a brute-force approach, which can estimate the utility of each possible combination such that the optimal one can be obtained. However, the formulated combinatorial optimization problems are usually NP-hard, thus the brute-force approach is not acceptable when there are a large number of workers or tasks. Therefore, existing research work commonly chooses to design approximate allocation algorithms (e.g., heuristic greedy, genetic algorithm, maximum flow algorithm, etc.) to achieve near-optimal solutions.

- **Incentive mechanism**. All the concerns mentioned in the "Participant-side factors" above may diminish a user's willingness to participate, which means that the participants must pay some price or suffer from certain risks when taking part in a CPSC task. Thus, how to design an incentive mechanism to compensate participants is a key issue in CPSC. For urban context sensing tasks, the cloud server can control participants through incentives that motivate them to provide the most accurate and useful reports. For urban context actuation tasks, the behavior of participants (e.g., driving, walking, reserving tables, etc.) are steered through incentives to achieve better resource utilization. Generally, incentives can be divided into two classes, money and enjoyment. More specifically, money is a participant's financial gain; enjoyment is the happiness she/he gains in the task participation, such as the interest in game play or his/her recognition by peers for his/her contributions.

- **Task scheduling and path design**. In many cases, one participant has to complete multiple CPSC tasks deployed in different locations. Thus, the participant should travel from one task location to another, which brings the optimal task scheduling or path design problem. For example, given a set of location-dependent CPSC tasks and a participant, the goal is to find a schedule which maximizes the number of tasks that he/she can complete while also considering travel cost and expiration time of the tasks. To tackle the scheduling problem of CPSC tasks, different approximation algorithms (e.g., dynamic programming, maximum flow algorithm, branch-and-bound algorithm, etc.) have been proposed with the consideration of traveling distance and the time to complete a task. Moreover, for real-world CPSC tasks, we should also take into account other factors (e.g., the road network, traffic status, etc.) when optimizing the task scheduling problem.

- **Location privacy preserving**. Privacy concern is one major human factor that may decrease participants' willingness to take part in CPSC tasks. In particular, location privacy is perhaps the most serious concern among all privacy issues, as CPSC tasks usually target at completing spatial tasks (sensing or actuation) within a city. A range of location privacy preserving techniques have been proposed for location-based services. For example, when using the cloaking protection method, the participant chooses a parameter of $l$ and then his/her uploaded location is a coarse-region covering $l$ fine-grained location cells; ε-differential-privacy bounds the adversary's posterior knowledge improvement over his/her prior knowledge of a user's location, while ε can be set by users' privacy preferences. Location privacy protection mechanisms generally include adding noises to participants' locations, which may lead to extra challenges for the assignment of CPSC tasks.

6. **Future Research Opportunities and Proposals**

In this section, we highlight the research gaps and future opportunities of CPSC, which may lead to novel solutions in this increasingly-important field.

- **Collaboration among different types of crowds**. The state-of-the-art studies on CPSC mainly focus on the collaboration among the same type of crowds, such as vehicle-to-vehicle, pedestrian-to-pedestrian, bicycle-to-bicycle, and drone-to-drone. However, relatively few studies have been conducted for the collaboration among different types of crowds, such as vehicle-to-pedestrian, vehicle-to-bicycle, drone-to-vehicle, drone-to-pedestrian, and so on. Using crowd steering as an example, we need to design a collaboration mechanism among vehicles, bicycles, and pedestrian to achieve an efficient ecosystem. Therefore, the study for collaboration among different types of crowds is a promising future research direction, which involves addressing several research issues, such as how to design a collaborative incentive model and how to define the overall system utility with the consideration of the heterogeneous crowds.

- **Combing opportunistic and participatory modes**. Existing CPSC solutions adopt either the opportunistic mode or the participatory mode. Motivated by the complementary nature of these two modes, there may be a hybrid solution, which can effectively integrate the opportunistic mode and the participatory mode. For example, we can recruit a number of participants (called opportunistic participants) to complete tasks during their routine trajectories. Then, we can further assign some other participants (called participatory participants) to locations where tasks cannot be completed by the opportunistic participants alone. The advantage of such hybrid solution is a better tradeoff between task completion ratio and cost. However, when the task assignment of these two types of participants is correlated (e.g., they share a total incentive budget), it is challenging to jointly optimize them, which remains a challenge for future research.

- **Combining crowd sensors and crowd actuators**. So far, the goal of existing CPSC applications can be classified as either context sensing or context actuation. However, for some application scenarios, we should integrate both the crowd sensors and crowd actuators. Using emergency event management (e.g., earthquake, terrorism attack) as an example, on the one hand, we need a number of crowd sensors to report current status through crowdsourced photo taking. On the other hand, we also need to guide the crowd actuators according to the current reported contexts, such as designing safe transferring routes or instructing people to help those in danger. In this case, as both the crowd sensors and crowd actuators work in the same location and time duration, it is challenging to make sure they collaborate effectively, which could be a future research opportunity.

- **Real-world deployment and evaluation**. So far, the evaluation of applications or frameworks in CPSC are mainly based on simulations, which is a common and major limitation. For urban sensing tasks, although some open real-world datasets on mobility traces of participants are used, other key factors, such as the number of tasks, the distribution of tasks, and sensor configurations, are commonly simulated by computer programs. For urban actuation tasks, the evaluation is more challenging, since it is difficult to evaluate the impact of certain strategies on the urban contexts with uncertainty in participants' behaviors. Therefore, to facilitate a more robust CPSC system, we need to spend more effort in the following two aspects: (1) conducting large-scale and real-world deployments and evaluations, and/or (2) improving parameterized simulations using data-driven approaches and behavioral models (such as utilizing more comprehensive user profiling information).

- **Social networks for the crowds**: There are some studies in CPSC focusing on how people collaborate as nodes in a social network. However, there could be social networks for other types of crowds (such as cars, bikes and mobile phones), and we could integrate the social networks among them. For example, Tom's car and his neighbor's car usually travel to the same places (e.g., a supermarket) and even with similar routes. The cars might find it useful to "follow" each other for sharing useful information (e.g., route congestion information). Here, the possible research issues may include: 1) how to build such social networks for different types of crowds, and 2) how to leverage the social networks among different types of crowds to enable interesting CPSC applications.

## 7. Conclusion

In this article, we present a survey of crowd-powered computing applications and techniques in smart cities. Specifically, by organizing state-of-the-art research in the perspective of context sensing and actuation, we present the concepts, applications, characteristics, research issues and techniques in this area, respectively. In the end, we point out the gaps with some future research opportunities and proposals.

**References**

1. Cocchia, A. (2014). Smart and digital city: A systematic literature review. In Smart city (pp. 13-43). Springer International Publishing.
2. R.K. Ganti, F. Ye, and H. Lei. Mobile crowdsensing: Current state and future challenges. IEEE Communications Magazine, 49:32–39, 2011.
3. Howe, Jeff. "The rise of crowdsourcing." Wired magazine 14.6 (2006): 1-4.
4. Zheng, Y., Capra, L., Wolfson, O., & Yang, H. (2014). Urban computing: concepts, methodologies, and applications. ACM Transactions on Intelligent Systems and Technology (TIST), 5(3), 38.
5. R. K. Rana, C. T. Chou, S. S. Kanhere, N. Bulusu, and W. Hu, "Ear-phone: an end-to-end participatory urban noise mapping system," in IPSN, 2010, pp. 105–116
6. S. Morishita, S. Maenaka, D. Nagata, M. Tamai, K. Yasumoto, T. Fukukura, and K. Sato, "Sakurasensor: quasi-realtime cherry-lined roads detection through participatory video sensing by cars," in Proceedings of the 2015 ACM International Joint Conference on Pervasive and Ubiquitous Computing. ACM, 2015, pp. 695–705.
7. Y. Zhu, Z. Li, H. Zhu, M. Li, and Q. Zhang, "A compressive sensing approach to urban traffic estimation with probe vehicles," IEEE Transactions on Mobile Computing, vol. 12, no. 11, pp. 2289–2302, 2013.
8. S. Mathur, T. Jin, N. Kasturirangan, J. Chandrasekaran, W. Xue, M. Gruteser, and W. Trappe, "ParkNet: drive-by sensing of road-side parking statistics," in Proceedings of the 8th international conference on Mobile systems, applications, and services. ACM, 2010, pp. 123–136.
9. K. K. Rachuri, C. Mascolo, M. Musolesi, and P. J. Rentfrow, "SociableSense: exploring the trade-offs of adaptive sampling and computation offloading for social sensing," in MobiCom, 2011, pp. 73–84
10. A. Singla, M. Santoni, G. Bart´ok, P. Mukerji, M. Meenen, and A. Krause. Incentivizing users for balancing bike sharing systems. In 29th AAAI Conference on Artificial Intelligence, 2015

11. Borean, C., Giannantonio, R., Mamei, M., Mana, D., Sassi, A., & Zambonelli, F. (2015, September). Urban crowd steering: An overview. In International Conference on Internet and Distributed Computing Systems (pp. 143-154). Springer International Publishing.
12. Sadilek, A., Krumm, J., & Horvitz, E. (2013). Crowdphysics: Planned and opportunistic crowdsourcing for physical tasks. SEA, 21(10,424), 125-620.
13. Tim Roughgarden. Twenty Lectures on Algorithmic Game Theory. Cambridge University Press, 2016.
14. https://www.justpark.com/ (crowd-based platform for sharing parking space).
15. https://www.opentable.com/start/home (crowd-based platform for sharing reserved tables).

**Authors' Bio**

Jiangtao Wang received his Ph.D. degree in Peking University, Beijing, China, in 2015. He is currently an assistant professor in Institute of Software, School of Electronics Engineering and Computer Science, Peking University. His research interest includes collaborative sensing, mobile computing, and ubiquitous computing.

Yasha Wang is a professor and associate director of National Research & Engineering Center of Software Engineering in Peking University, China. His research interest includes urban data analytics and ubiquitous computing. He has published more than 50 papers in prestigious conferences and journals, such as ICWS, UbiComp, ICSP and etc.

Daqing Zhang is a professor at Peking University, China, and Télécom SudParis, France. His research interests include context-aware computing, urban computing, mobile computing, and so on. He served as the General or Program Chair for more than 10 international conferences. He is an Associate Editor for ACM Transactions on Intelligent Systems and Technology, IEEE Transactions on Big Data, and others.

Qin Lv is an associate professor in the Department of Computer Science, University of Colorado Boulder. Her main research interests are data-driven scientific discovery and ubiquitous computing. Lv is an associate editor of ACM IMWUT, and has served on the technical program committee and organizing committee of many conferences.

Chao Chen is an associate professor at College of Computer Science, Chongqing University, Chongqing, China. His research interests include pervasive computing, urban logistics, data mining from large-scale taxi data, and big data analytics for smart cities.